\documentclass[english,preprint,showpacs,preprintnumbers,amsmath,amssymb,showkeys]{revtex4}

\usepackage[T1]{fontenc}
\usepackage[latin9]{inputenc}
\usepackage{color}
\usepackage{units}
\usepackage{amssymb}
\usepackage{graphicx}
\usepackage{esint}
\usepackage{bm}
\usepackage{natbib}
\usepackage{amsmath}
\usepackage{chemformula} 
\usepackage[colorlinks=true, citecolor=blue]{hyperref}
\usepackage{pdfpages}
\usepackage{ulem}
\setcitestyle{journalcolor= blue}

\usepackage{babel}

\begin{document}
	\title{Electric field tunable valley-Zeeman effect in bilayer graphene heterostructure: realization of the spin-orbit  valve effect}
	
	\author{Priya Tiwari}
	\affiliation{Department of Physics, Indian Institute of Science, Bangalore 560012, India}
	\author{Saurabh Kumar Srivastav}
	\affiliation{Department of Physics, Indian Institute of Science, Bangalore 560012, India}
	
	\author{Aveek Bid}
	\email{aveek@iisc.ac.in}
	\affiliation{Department of Physics, Indian Institute of Science, Bangalore 560012, India}

	\begin{abstract}
		
			We report the discovery of electric field-induced transition from a topologically trivial to a topologically non-trivial band-structure in an atomically sharp heterostructure of bilayer graphene (BLG) and single-layer \ch{WSe2} per the theoretical predictions of by Gmitra et al. [\textbf{Phys. Rev. Lett. 119, 146401 (2017)}]. Through detailed studies of the quantum correction to the conductance in the BLG, we establish that the band-structure evolution arises from an interplay between proximity induced strong spin-orbit interaction (SOI) and the layer-polarizability in BLG. The low-energy carriers in the BLG experience an effective valley-Zeeman SOI that is completely gate-tunable to the extent that it can be switched on/off by applying a transverse displacement field or can be controllably transferred between the valence and the conduction band. We demonstrate that this results in the evolution from weak localization to weak anti-localization at a  constant electronic density as the net displacement field is tuned from a  positive to a negative value with a concomitant SOI-induced splitting of the low-energy bands of the BLG near the $K$($K'$)--valley which is a unique signature of the theoretically predicted spin-orbit valve effect.  Our analysis shows that quantum correction to the Drude conductance in Dirac materials with strong induced SOI can only be explained satisfactorily by a theory that accounts for the SOI induced spin-splitting of the BLG low-energy bands. Our results demonstrate the potential for achieving highly tunable devices based on the valley-Zeeman effect in dual-gated  2-dimensional materials.
\end{abstract}

\maketitle
Over the last decade, the search for new quantum materials has attracted much interest from physicists and material scientists. A route to engineer new materials with desirable set of properties is via van der Waals (vdW) heterostructures of two-dimensional (2D) materials~\cite{Liu2016,Geim2013,doi:10.1146/annurev-matsci-070214-020934,C4NR03435J,doi:10.1002/adfm.201706587, Novoselovaac9439,kumar2018localization}. This has led to the emergence of materials with designer properties not present in the parent components. For example, there is a particular interest in gaining the ability  to manipulate the spin-degree of freedom in graphene electrically. The intrinsic SOI in graphene being very small (10$\mu$eV)~\cite{gmitra2009band}, this endeavor has met with minimal success. Several efforts have been made to increase the SOI in graphene, including doping it with the heavy atoms~\cite{weeks2011engineering} or topological nanoparticles~\cite{hatsuda2018evidence}; this has been achieved at the cost of degrading the quality of graphene. 

Recently, both theoretical~\cite{PhysRevLett.119.146401, doi:10.1021/acs.nanolett.7b03604,PhysRevLett.119.206601,PhysRevB.92.155403,PhysRevLett.119.196801} and experimental~\cite{PhysRevLett.121.127702,island2019spin, PhysRevX.6.041020, Benitez2018, PhysRevB.97.075434, PhysRevB.97.045414, PhysRevLett.121.127703} studies have indicated that strong SOI can be `induced' in graphene when placed on ultra-thin films of transition metal dichalcogenides (TMD).  Broken inversion symmetry, coupled with the presence of heavy atoms, leads to the appearance of strong intrinsic Ising spin-orbit coupling in single-layer TMD~\cite{PhysRevLett.119.146401, doi:10.1021/acs.nanolett.7b03604}. When placed in close proximity to graphene, the hybridization of d-orbitals of the TMD with those of carbon in graphene leads to the appearance of a large SOI in the latter. The case of BLG is particularly interesting: As the SOI is induced by a modification of the band structure of BLG, it opens up an avenue for preparing materials imbibing strong spin-polarizing SOI along with other desirable properties of BLG like high-mobility, gate-tunable bandgap, and valley/layer polarizability for possible spintronics, and optoelectronics applications~\cite{C7CS00864C}.

This proximity-induced SOI in the low-energy bands of BLG has two primary components -- a valley-Zeeman term that causes spin-splitting of the band structure and a Kane-Mele term that opens up a topological gap at the two non-equivalent valleys, $\pm~K$~\cite{PhysRevB.99.205407,doi:10.1021/acs.nanolett.7b03604, PhysRevLett.119.146401,PhysRevX.6.041020, C7CS00864C}. It was predicted that for BLG on a single-layer of \ch{WSe2}, the strength of the SOI in the valence band (VB) of the BLG ($\sim$2~meV) could be two orders of magnitude larger than that in the conduction band (CB)~\cite{PhysRevLett.119.146401}. This is because the VB  is formed by the non-dimer carbon atom orbitals in the BLG bottom layer, which are in proximity with \ch{WSe2}. In contrast, the CB is formed by the non-dimer orbitals of the top layer of the BLG, where the SOI strength remains similar to that in intrinsic BLG~\cite{PhysRevLett.119.146401, Wang2015}. The induced SOI is manifested as a relatively large, spin-polarized splitting of the VB of the BLG. 

Further, by applying an electric field of sufficient strength perpendicular to the interface, it is possible to invert the band structure, i.e., close the bandgap and reopen it - but this time transferring the SOI induced splitting to the conduction band. Calculations show that by tuning the electric field, it is thus possible to transform the band-structure of the BLG system from a topologically trivial to non-trivial.   In recent work, we have demonstrated that this leads to the appearance of the time-reversal invariant $\mathbb{Z}_2$ topological phase in BLG~\cite{tiwari2020observation}. Although SOI induced splitting of the VB/CB   was experimentally verified recently using penetration field capacitance measurements~\cite{island2019spin}, direct transport observation of electric field-induced band-inversion is missing.

In this letter, we present results of magneto-transport measurements in dual gated hBN/BLG/single-layer \ch{WSe2}/hBN heterostructures explicitly designed to explore the proposal of  Gmitra et al.~\cite{PhysRevLett.119.146401}. We show a precise electric field tuning of the band structure topology of BLG -- from a doubly-split CB and an un-split VB  at $D>0$ to a doubly-split VB and an un-split CB at $D<0$.  Thus, the band structure splitting could be controllably transferred from the VB to the CB using the electric field as an external tuning parameter. In a given band, the strength of the SOI was tunable by the electric field from a significant value to negligibly small, thus realizing the spin-orbit valve~\cite{PhysRevLett.119.146401}. 

We further establish out that the conventional McCann-Fal'ko (MF) equation~\cite{PhysRevLett.108.166606}, which has been used almost universally to study W(A)L in graphene/TMD heterostructures, is not appropriate to quantify the weak localization physics when there is an SOI induced spin-splitting of bands, as in our case. Instead, we show that the models developed by S. Ilic et al.~\cite{PhysRevB.99.205407} and Ochoa et al.~\cite{PhysRevB.90.235429} can account for the scattering mechanisms of the spin-split bands. We found that in the WAL case, the relevant parameters determining magneto-transport are the phase coherence time $\tau _{\phi}$ and the anti-symmetric spin relaxation $\tau _{asy}$ (which breaks the $z\longrightarrow -z$ inversion symmetry). On the other hand, in the WL regime where the induced SOI is very small, the relevant time scales are $\tau _{\phi}$ and intervalley scattering time $\tau _{iv}$. 

\begin{figure*}
	\includegraphics[width=0.75\textwidth]{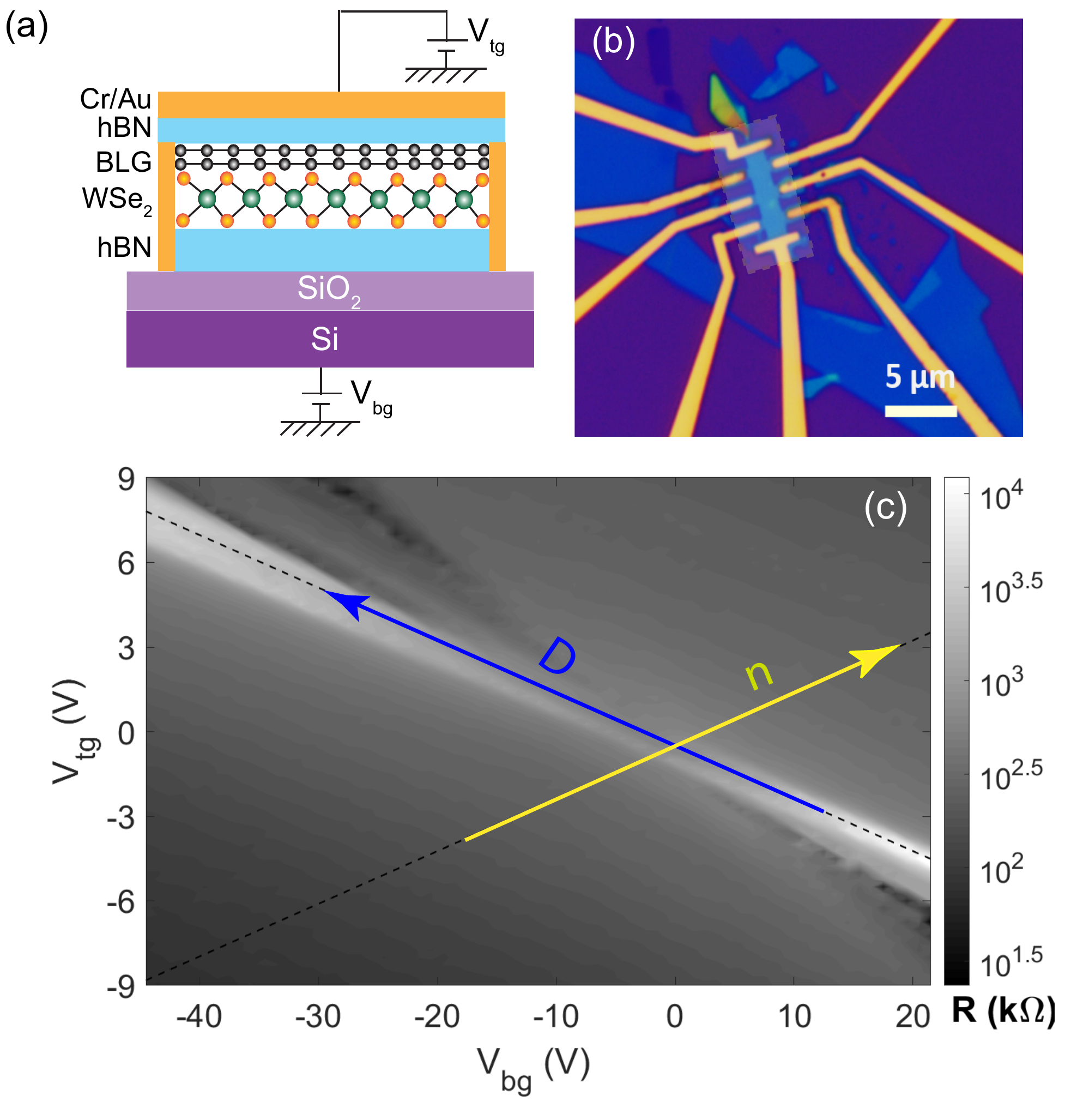}
	\small{\caption{(a) Schematic of the device structure. The BLG and single-layer \ch{WSe2} stack is sandwiched between two hBN flakes of thickness $\sim 20$~nm. (b) Optical image of the device without top-gate structure -- the gray rectangle shows the position of the  top gate. (c) Contour plot of four-probe resistance measured in the  $V_{bg}$-$V_{tg}$ plane on logarithmic scale. The blue and yellow arrows indicate the  directions of increasing $D$ and $n$, respectively.
			\label{fig:figure1}}}
\end{figure*}

\begin{figure*}
	\begin{center}
		\includegraphics[width=1.0\textwidth]{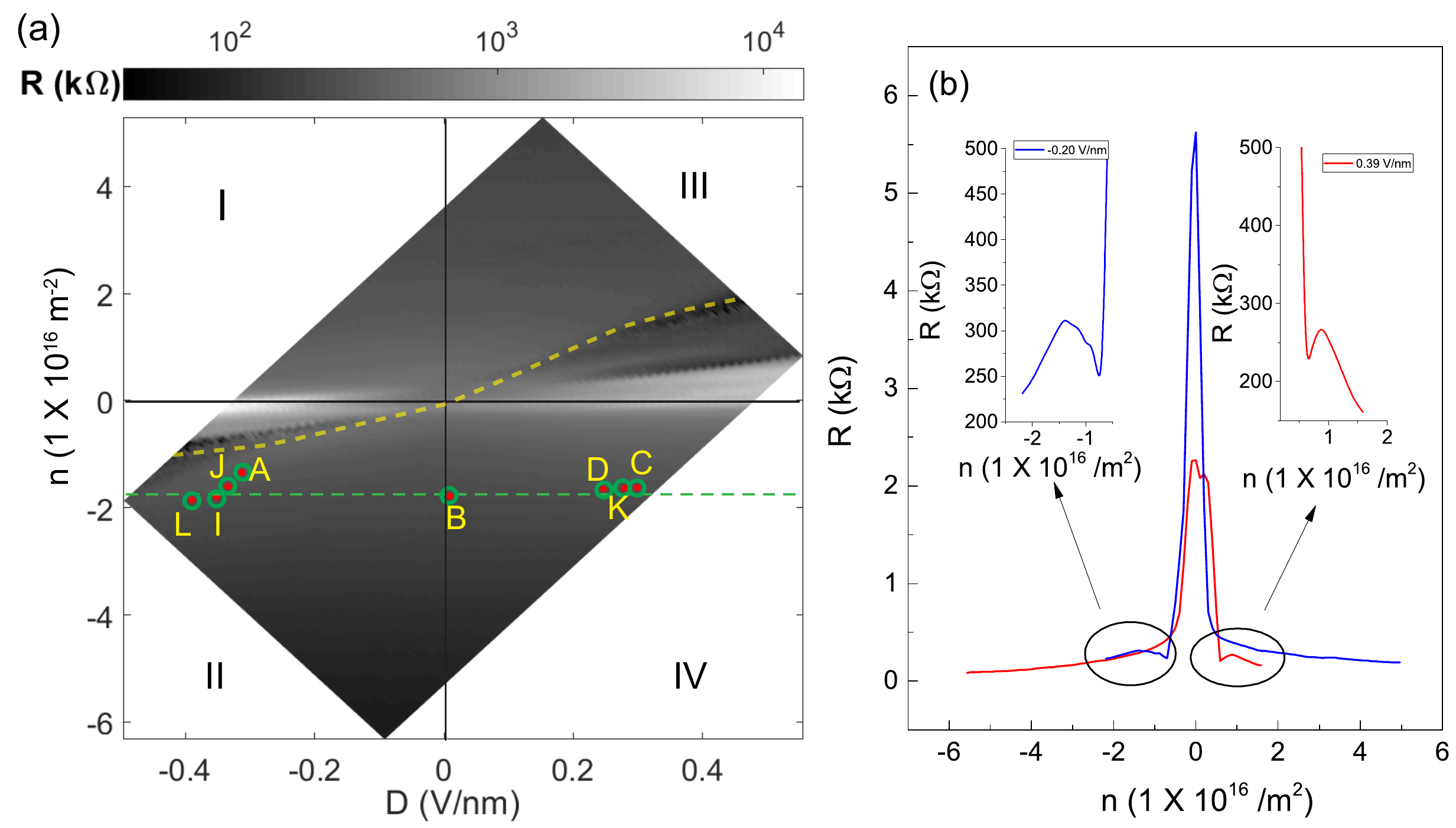}
		\small{\caption{(a) A 2-dimensional contour plot of $R$ as a function of $n$ and $D$. The asymmetric feature around the primary Dirac point ( marked by the dashed line) is a consequence of the fact that only one of the bands (VB for negative $D$ and CB for positive $D$) get split due to induced SOI~\cite{PhysRevLett.119.146401}. The data were acquired at $T=20$~mK and $B=0$~T.   The labels I-IV mark the four quadrants in the $n$-$D$ plane. The corresponding regimes are marked in the two schematic band structures in Fig.~\ref{fig:figure3}.  The W(A)L measurements reported in this letter were performed along the green line (some of the points are labeled). (b) Plots of $R$ versus $n$ for $D = -0.2$~V/nm (blue line) and $D=0.4$~V/nm (red line). The insets show the data zoomed in around the secondary peaks (marked in the main panel by dashed ovals) arising from SOC-induced splitting of the bands.    				\label{fig:figure2}}}
	\end{center}
\end{figure*}

Heterostructures of hBN/BLG/single-layer \ch{WSe2}/hBN were fabricated using standard dry transfer technique~\cite{pizzocchero2016hot, Wang614} (see Supplemental Material \cite{supplement}) 1). Single-layer \ch{WSe2} was used as it induces a much stronger SOI in BLG  than thicker flakes. Fig.~\ref{fig:figure1}(a) is the schematic showing the dual-gated architecture of the devices. Fig.~\ref{fig:figure1}(b) is optical images of a device without top-gate. All electrical measurements were performed at $T$ = 20~mK in a dilution refrigerator unless specified, using standard low-frequency ac lock-in techniques.

The presence of both top- and bottom-gates in the device allowed independent tuning of $n$ and  $D$ via the relations $n=((C_{tg}V_{tg}+C_{bg}V_{bg})/e)-n_0$ and $D=((C_{tg}V_{tg}-C_{bg}V_{bg})/2\epsilon_{o})-D_0$. Here, $n_0$ is the residual charge-density due to doping and $D_0$ is net internal displacement field. Fig.~\ref{fig:figure1}(c) shows variation of the four-probe longitudinal resistance, $R$ with $V_{tg}$ and $V_{bg}$ on a logarithmic scale. The blue and yellow arrows indicate the directions of increasing $D$ and $n$, respectively. A contour plot of the logarithm of the 4-probe longitudinal resistance $R$ as functions of  $n$ and $D$ ((Fig.~\ref{fig:figure2}(a)) shows that the resistance at the Dirac point (DP) increases with an increase in $\left|D\right|$ establishing the opening of a bandgap in BLG. Along with the prominent peak at DP, an asymmetric feature is observed near it as outlined by the yellow dashed curve in Fig.~\ref{fig:figure2}(a). This is a direct consequence of the predicted spin-splitting of the VB and CB due to proximity induced SOI in BLG~\cite{PhysRevLett.119.146401} and observed recently through penetration field capacitance studies~\cite{island2019spin} and transport measurements~\cite{tiwari2020observation}.This asymmetric feature in the $n-D$ plane depends on the direction and magnitude of displacement field - appearing in the positive (negative) $n$ regime for positive (negative) $D$ (Fig.~\ref{fig:figure2}(b)).

\begin{figure*}
	\begin{center}
		\includegraphics[width=0.75\textwidth]{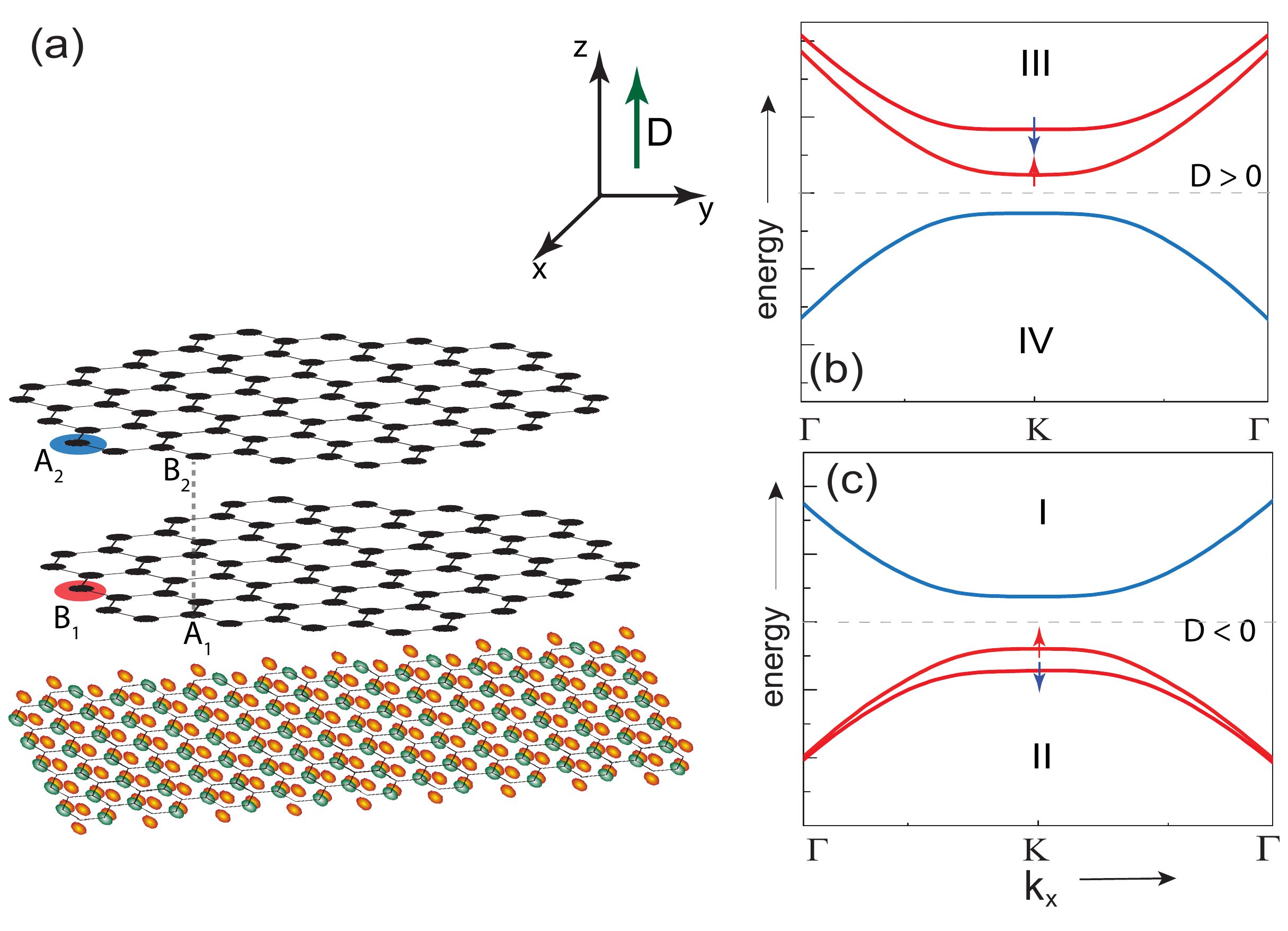}
		\small{\caption{(a) Schematic showing the single-layer \ch{WSe2} and the two layers of the BLG. The low-energy bands in the BLG arise from hopping between the orbitals on the non-dimer pair of sites formed of ${B_1}$ sub-lattice of the lower layer (one such atom is marked in red) and the ${A_2}$ sub-lattice of the top layer (one such atom is marked in blue)~\cite{McCann_2013}. The system lies in the $x-y$ plane, while the positive direction of $D$ (indicated by the green arrow) is the positive $z$-axis. (b) Schematic of the bulk band structure near the $K$-valley for positive values of $D$. The electrons localized in the bottom-layer ($B_1$ orbital) of the BLG form the spin-split CB, and those in the top-layer ($A_2$ orbital) of the BLG form the VB~\cite{PhysRevLett.119.146401}.  (c) Schematic of the bulk band structure plotted near the $K$-valley for negative $D$. The electrons localized in the $B_1$ orbital of the BLG now form the spin-split VB, and those in the $A_2$ orbital of the BLG form the CB. The BLG thus undergoes a band-inversion as $D$ changes sign~\cite{PhysRevLett.119.146401}. \label{fig:figure3}}}
	\end{center}
\end{figure*}

These observations can be understood by noting that the presence of the single-layer \ch{WSe2} causes the BLG to experience, in addition to the strong valley-Zeeman SOI, a displacement field of the order of -0.25 V/nm making it layer-polarized. For $D<0$, the Bloch waves associated with the VB and CB of the BLG reside on the bottom-layer and top-layer, respectively (Fig.~\ref{fig:figure3}(a)). Since the bottom layer is in close contact with the \ch{WSe2}, the VB is prominently spin-split. This splitting increases as $D$ is made more negative via the gate-voltages.  Upon changing $D$ to increasingly positive the bandgap initially closes as the external field cancels out that due to the \ch{WSe2}. On increasing $D$ further, the bandgap reopens, but now the layer-polarization of the BLG reverses direction, causing the lower-layer to form the spin-split CB (Fig.~\ref{fig:figure3}(b)). Calculations show that this will cause the system to undergo a displacement field induced topological phase transition into the anomalous valley Hall phase~\cite{doi:10.1021/acs.nanolett.7b03604} and quantum spin hall phase~\cite{tiwari2020observation} as recently demonstrated. 

Having established displacement field induced band-inversion in our device, we move on to probe the electron dynamics as we transit from the trivial VB at $D>0$ to the non-trivial spin-split VB at $D<0$. The SOI mechanism can be effectively probed by studying the quantum correction to the conductance in the presence of a perpendicular magnetic field $B$. The magnetoconductance measurements were performed in the parameter space of $D$-$n$ plane along the green line shown in Fig.~\ref{fig:figure2}(b) for a $n=-1.8 (\pm 0.05)\times10^{16}$~m$^{-2}$ while changing |D| from negative to positive. To avoid contamination of the data by universal conductance fluctuation~\cite{lee1985universal},  each plot is an ensemble average of 40 traces measured at very close values of $n$  (spanning the range $\Delta n=\pm0.05\times10^{16}$~m$^{-2}$).  

The four-terminal magnetoconductance  $\Delta \sigma (B) = \sigma (B)- \sigma (0)$ traces plotted in Fig.~\ref{fig:figure4}(a) show a clear crossover from WAL to WL as the displacement field changes from negative to positive. This transition can be understood as follows: for negative $D$ and chemical potential  $E_F$ lying in VB (\textbf{regime II} in Fig.~\ref{fig:figure2} and Fig.~\ref{fig:figure3}), the holes experience a strong  SOI. This, along with the breaking of the {$z\ensuremath{\rightarrow}\ensuremath{-}z$ symmetry, leads to the observed WAL~\cite{PhysRevLett.108.166606}. As the displacement field becomes positive, the electrons in conduction bands experience strong SOI, while the SOI in VB reduces to a value intrinsic to BLG, which, as discussed before, is extremely small. Since the $E_F$ still lies in the VB (\textbf{regime IV} in Fig.~\ref{fig:figure2} and Fig.~\ref{fig:figure3}), we observe WL, which is the characteristic of BLG due to its intrinsic Berry phase of 2$\pi$.  This transformation of the magnetoconductance from WAL to WL again demonstrates displacement field induced band-inversion in BLG/\ch{WSe2} heterostructures. Fig.~\ref{fig:figure4}(a) also shows the magnetoconductance for $D~\sim 0.01$~V/nm (blue data points). In this cross-over regime, the orbitals localized in both the upper- and the lower-layer of the BLG contribute equally to the valance and conduction bands leading to a magnetoconductance curve, which is a complex admixture of WL (arising from the intrinsic Berry curvature of unbiased BLG) and WAL (due to the induced SOC).

\begin{figure*}
	\begin{center}
		\includegraphics[width= 1.0\textwidth]{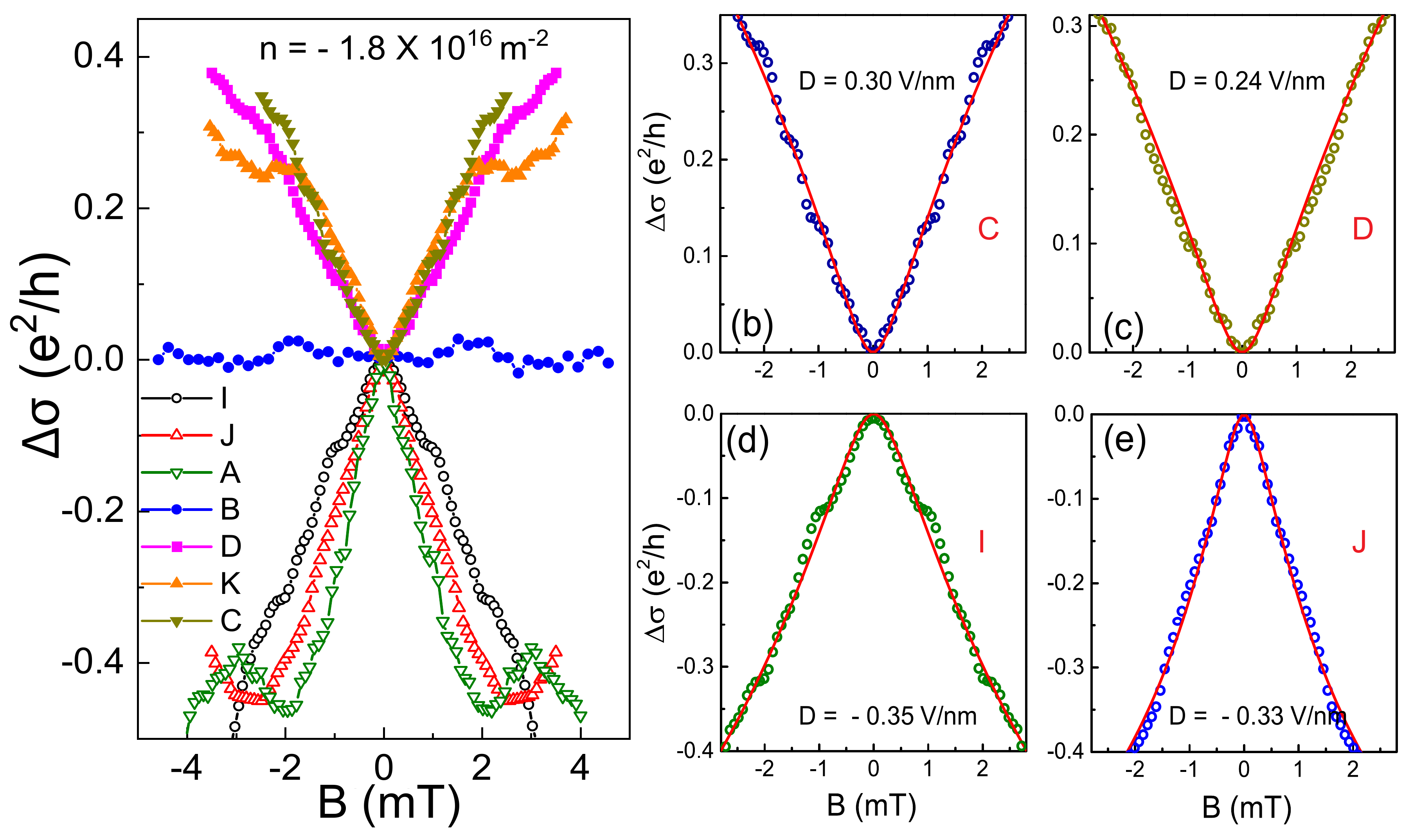}
		\small{\caption{(a) Plots showing the dependence of the ensemble averaged  magnetoconductance data on  $B$ for $n = -1.8\times10^{16}$~m$^{-2}$ at different $ D$. The data were collected along the green dotted line in Fig.~\ref{fig:figure2}. For $D < 0$,  WAL is observed which changes to WL as the $D$ is made sufficiently positive. The blue data points (measured at point B in Fig.~\ref{fig:figure2}) show the magnetoconductance near $D~\sim$~0.01 V/nm. (b, c) The open circles are the ensemble averaged WL data taken for $D > 0$ and $n < 0$ (\textbf{regime IV} in Fig.~\ref{fig:figure2} and Fig.~\ref{fig:figure3}). The solid red lines are the fits to the data using Eqn.~\ref{1}. (d, e) The open circles are the ensemble averaged WAL data taken for $D < 0$ and $n < 0$ (\textbf{regime II} in Fig.~\ref{fig:figure2} and Fig.~\ref{fig:figure3}). The solid red lines are the fits to the data using Eqn.~\ref{2}. The letters C, D, I and J refer to the points in the $n-D$ plane in Fig.~\ref{fig:figure2} at which the data were collected.
				\label{fig:figure4}}}
	\end{center}
\end{figure*}

As mentioned earlier, the MF equation cannot be used to extract the scattering time-scales in our system, as it does not take into account the spin splitting of bands. To get a quantitative estimate of different scattering time scales involved, the magnetoconductance data was analyzed following the approach recently~\cite{PhysRevB.99.205407} which discusses the two different regimes taking into account  the relative strengths of the SOI interaction, $\Delta_{so}$~\cite{PhysRevB.97.075434} and inter-valley scattering rate, $\tau_{iv}^{-1}$. The Valley-Zeeman SOI is a strong source of symmetric spin relaxation rate~\cite{PhysRevB.97.075434,garcia2018spin}. In the presence of this component of SOI, the net symmetric spin relaxation rate becomes ${\tau}_{sym}^{-1}\rightarrow\tau_{sym}^{-1}+2\Delta_{so}^{2}\tau_{iv}$. The asymmetric spin relaxation rate $\tau_{asym}^{-1}$, on the other hand, arises from the Rashba SOI. When $E_F$ lies in VB and the CB is spin-split ($D > 0$), this regime is characterized by $\tau_{iv}^{-1}> \Delta_{so}$. In this WL regime, for phase coherence rate $\tau_{\phi}^{-1}$ comparable to $\tau_{iv}^{-1}$,  the magnetoconductance can be analyzed using the following equation~\cite{PhysRevB.99.205407}:
\begin{align}
	\Delta \sigma (B) = \frac{e^2}{\pi h} \Huge{[} F\left(\frac{B}{B_{\phi}}\right)- F\left(\frac{B}{B_{\phi}+2B_{iv}}\right) +2F\left(\frac{B}{B_{\phi}+B_{iv}}\right)\Huge{]} 
	\label{1}
\end{align}
where $F (z) = ln(z)+\psi \left( \frac{1}{2} +\frac{1}{z}\right)$ , $\psi$ is the digamma function, $ B_{i} = \hbar/(4e\tilde{D}\tau_{i})$ and $\tilde{D}$ is the diffusion coefficient. For $D < 0$, and $E_F$ lying in the spin-split VB, the system is characterized by $\Delta_{so}>\tau_{iv}^{-1}$. Due to strong valley-Zeeman SOI, WL gets suppressed, and one will observe WAL described by the following equation~\cite{PhysRevB.99.205407}:   
\begin{equation}
	\Delta \sigma (B) = \frac{e^2}{2\pi h} \left[ F\left(\frac{B}{B_{\phi}+2B_{asy}}\right)-F\left(\frac{B}{B_{\phi}}\right) \right] 
	\label{2}
\end{equation}
where $B_{asy}$ corresponds to the scattering rate $\tau_{asy}^{-1}$. Physically $\tau_{asy}^{-1}$ corresponds to the spin-flip processes that break the $z\longrightarrow-z$ symmetry induced by Rashba SOI.

Figs.~\ref{fig:figure4}(b,c) are plots of WL and the data were fitted using Eqn.~\ref{1}. The WAL data are shown in figs.~\ref{fig:figure4}(d,e) and have been fitted using Eqn.~\ref{2}. The extracted time-scales are tabulated in Supplemental Material \cite{supplement}). In the case of WAL, $\tau_{asy}$ was found to be $\sim 0.5-2$~ps, smaller than $\tau_{\phi}$, which varies between 7-22~ps, consistent with the earlier  observations~\cite{PhysRevB.97.045414, PhysRevX.6.041020} and theoretical predictions~\cite{C7CS00864C}. For the WL case, $\tau_{iv}$ (0.9-2~ps) was found to be smaller than $\tau_{\phi}$ (9-14~ps). The above analysis establishes that electron interference effects are dominated by spin-flip processes in the spin-split band and by inter-valley scattering in the spin-degenerate bands of BLG. It also shows that the strength of the SOI can be tuned from a significant value to negligibly small by the external electric field -- this unequivocally establishes the realization of the spin-orbit valve effect and is one of the central results of this letter~\cite{PhysRevLett.119.146401}.

\begin{figure*}	
	\begin{center}
		\includegraphics[width= 0.80\textwidth]{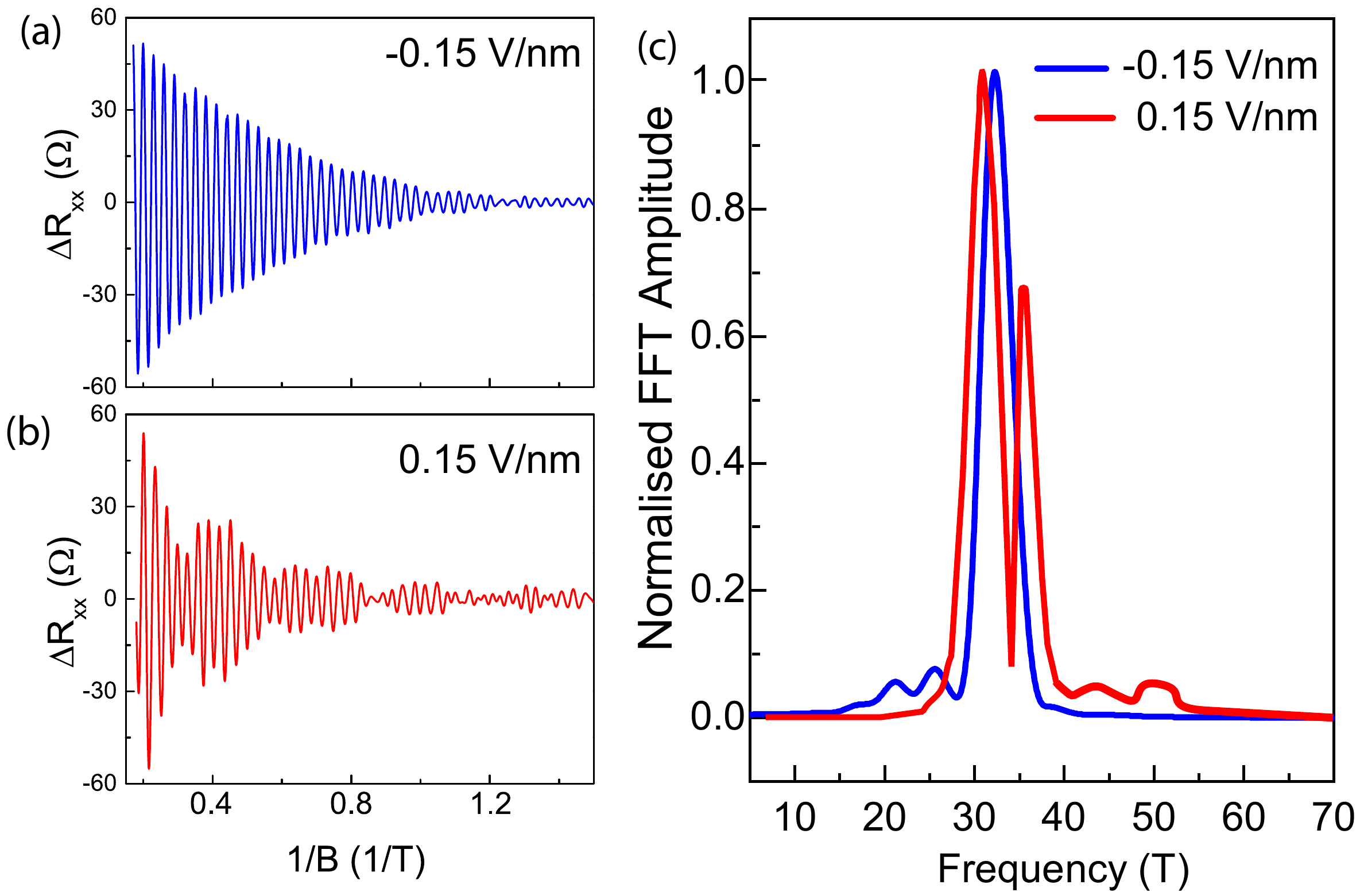}
		\small{\caption{Shubnikov-de Haas oscillations (measured at $n\sim3 \times10^{16}$~m$^{-2}$) plotted versus $1/B$  for (a) $D\sim-0.15$~V/nm and (b) $D\sim0.15$~V/nm. The beating in the oscillations observed for positive  values of $D$ indicate the presence of two close-by frequencies. (c) The red and blue curves show respectively the Fourier transform of the SdH oscillations for $D\sim0.15$~V/nm and $D\sim-0.15$~V/nm. One can see a clear change in the Fermi surface topology from single (\textbf{regime I} in Fig.~\ref{fig:figure2} and Fig.~\ref{fig:figure3}) to doubly split (\textbf{regime III} in Fig.~\ref{fig:figure2} and Fig.~\ref{fig:figure3}) on changing $D$ from negative to positive. 
				\label{fig:figure5}}}
	\end{center}
\end{figure*}

Additional support to our claim of electric field tuned band-inversion and SOI induced band-splitting comes from the results of  Shubnikov-de Haas (SdH) oscillations measurements. The data were collected in a similar device of higher mobility ({$\sim$1,10,000 $\mathrm{cm^2V^{-1}s^{-1}}$}) at a constant number density  $n\sim3\times10^{16}$~m$^{-2}$. Each measurement consisted of $\sim$3000 data points and was performed for several values of $D$ between -0.15~V/nm and 0.15~V/nm. For $n>0$ and $D>0$ (\textbf{regime III}) we observe a beating pattern in the SdH oscillations -- fig.~\ref{fig:figure5}(b) shows a representative data for $D=0.15$~V/nm. The beating results from the presence of two close-by frequencies of nearly equal magnitudes in the signal. The FFT data plotted in Fig.~\ref{fig:figure5}(c) confirms the presence of two frequencies at $\sim$~31.5~T and $\sim$~36.1~T  reflecting the presence of two extrema of the Fermi surface at this energy. This SOI-induced band-splitting has been reported previously in BLG/\ch{WSe2} heterostructures~\cite{PhysRevX.6.041020}  -- although in that case, a study of displacement field tunability was lacking. When the experiment was repeated for $D<0$ (\textbf{regime I}) no beating pattern was observed (see fig.~\ref{fig:figure5}(a) for data at $D= -0.15$~V/nm) -- the FFT of the data has a single peak at $\sim$~32.6~T (fig.~\ref{fig:figure5}(c)). Thus, a clear transition from a single Fermi surface to two Fermi surfaces is observed as $D$ is varied from negative {\textbf{(regime I)}} to positive {\textbf{(regime III)}} when $E_F$ lies in the CB. The results of SdH oscillations complement our observation, from low-field magnetoresistance measurements, of transition from a single Fermi surface to two Fermi surfaces as $D$ is varied from positive {\textbf{(regime IV)}} to negative {\textbf{(regime II)}} for $E_F$ lying in the VB. We thus demonstrate `on-demand' electric field induced band structure modulation of BLG over all four regimes marked in Fig.~\ref{fig:figure2}(a).


In summary, we demonstrate an electric field induced band structure engineering in BLG. Through magneto-transport measurements, we observe a transition from a topologically trivial to a non-trivial band in BLG. The low-energy carriers in the BLG experience an effective valley-Zeeman SOI that is gate-tunable to the extent that it can be switched on/off by applying a transverse displacement field or can be controllably transferred between the valence and the conduction band. This leads to the realization of the predicted spin-orbit valve effect.


\begin{acknowledgments}
	The authors acknowledge fruitful discussions with Tanmoy Das and Sujay Ray and facilities in CeNSE, IISc. S.K.S. acknowledges financial support from PMRF, MHRD. AB acknowledges funding from SERB (HRR/2015/000017) and DST (DST/SJF/PSA-01/2016-17). 
\end{acknowledgments}

\section*{References}
\bibliography{qshe}	


\newpage
\thispagestyle{empty}
\mbox{}


\section*{Supplementary material}

# S1. DEVICE FABRICATION AND CHARACTERIZATION

The device consists of dual gated bilayer graphene/WSe$_2$ heterostructure fabricated using standard dry transfer technique [1, 2]. The process involves mechanical exfoliation of graphite, WSe$_2$, and bulk hBN crystal on SiO$_2$/Si wafer to obtain bilayer graphene (BLG), single layer WSe$_2$ and thin hBN of thickness 25 to 40 nm. BLG and single layer WSe$_2$ were identified using the optical microscope and confirmed by room temperature photoluminescence (PL) and Raman spectroscopy – the data are presented in Fig. S1(a) and (b) respectively. The observation of high intensity peak at ∼1.65 eV in the PL spectra is due to the direct bandgap of single-layer WSe$_2$. Furthermore, in the Raman spectra a peak at ∼ 250 cm$^{-1}$ was observed; the absence of B$_{2g}^1$ peak at ∼ 307 cm$^{-1}$ confirms the single layer WSe$_2$. The other two peaks observed at 1580 cm$^{-1}$ (G peak) and 2800 cm$^{-1}$ (2D peak) are of graphene family. Peak ratio and spectral decomposition of 2D peak into four Lorentzians shown in Fig. S1(c) confirms bilayer nature of graphene.

The fabrication of the desired heterostructure involved the following steps: we made a Poly-Bisphenol-A-Carbonate (PC) coated Polydimethyl-siloxane (PDMS) block mounted on a glass slide attached to tip of a custom-built micro-manipulator under a microscope to pick-up the exfoliated flakes. Pickup and transfer of final stack were done at 90°C and 180°C, respectively. First the hBN flake was picked up and was aligned over the BLG. Similarly all the other flakes were picked up and aligned in the desired sequence and the final stack consisting of hBN/BLG/WSe2/hBN was transferred on top of an oxidized silicon wafer at 180°C. The prepared stack was then cleaned in chloroform (CHCL$_3$) followed by acetone and isopropyl alcohol.

This was followed by standard electron beam lithography technique to define the edge contacts. The edge contact was achieved by the reactive ion etching (mixture of CHF$_3$ and O$_2$ gas were used with flow rate of 40 sccm and 4 sscm, respectively at 25°C with RF power of 60W). Finally, a deposition of Cr/Au (5/60 nm) followed by lift-off in hot acetone and isopropyl alcohol created the electrical contacts. Top gate was patterned using electron beam lithography followed by Cr/Au deposition.

To calculate the mobility ($\mu$) of the device, we plotted conductivity ($\sigma$) with the number density ($n$) shown in Fig. S1(d). The slope of a linear fit to the data gave the value of $e\mu$. For the first device described in the main text (device A), the mobility was relatively low

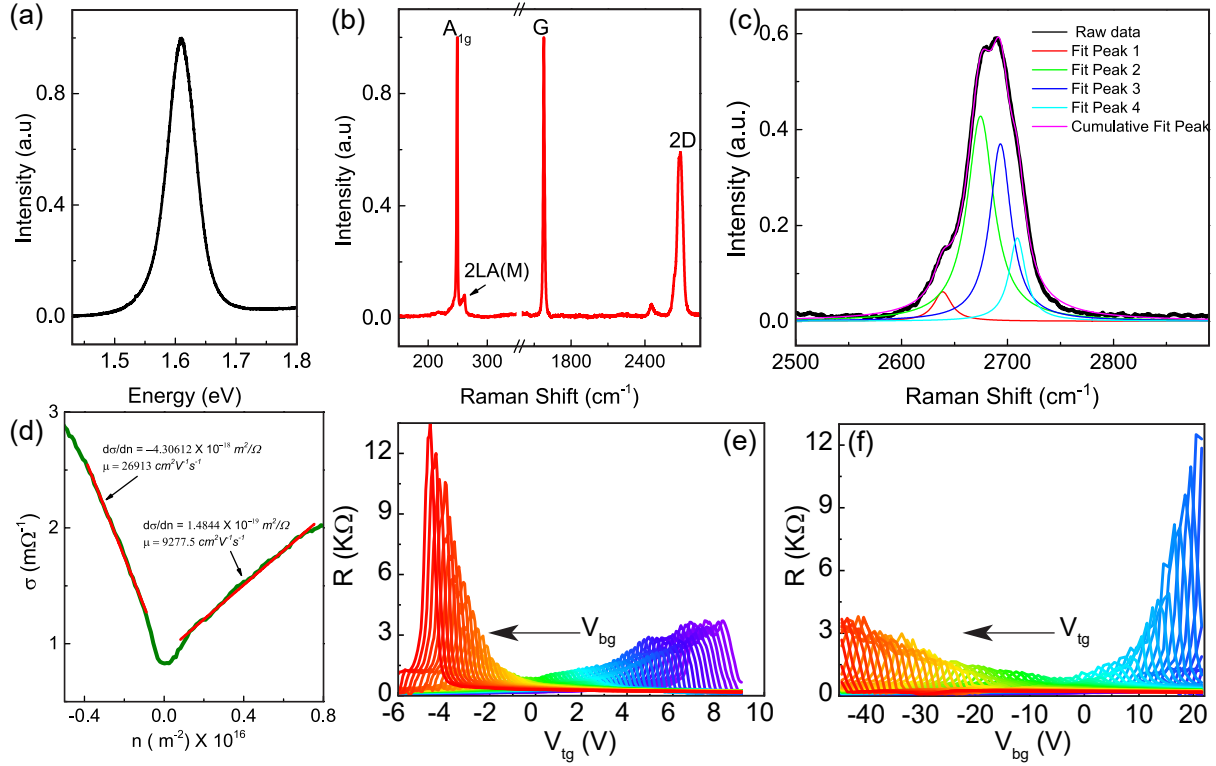

Figure S1: (a) Photoluminescence response of WSe$_2$ flake at room temperature, the peak at 1.65 eV establishes single-layer nature of the flake. (b) Room temperature Raman spectra of the WSe$_2$ and BLG flakes. The peaks intensities of single-layer WSe$_2$ and BLG are marked. (c) 2D peak of BLG is fitted with multiple Lorentzian curve showing the spectral decomposition into the four Lorentzian components. (d) Plot of four-probe conductance versus $n$ at 20 mK, the solid red line represents the linear fit to equation $\sigma = ne\mu$. (e) Plots of four-probe resistance $R$ versus $V_{tg}$ at different values of $V_{bg}$ – the arrow represents the increasing magnitude of $V_{bg}$. (f) Corresponding plots of $R$ versus $V_{bg}$ at different values of $V_{tg}$ – the arrow represents the increasing magnitude of $V_{tg}$. The data were collected at $T$=20 mK.

– the system thus obtained was in the diffusive regime ($l << l_\phi$) and suitable for studying localization physics. For the device B, the mobility was higher making it suitable for studying SdH oscillations. In Fig. S1(e), the four-probe resistance $R$ is plotted as a function of top gate voltage $V_{tg}$ for different values of the back gate voltage $V_{bg}$. The corresponding plots of $R$ versus $V_{bg}$ at varying $V_{tg}$ are shown in in Fig. S1(f). As expected, in both cases, the charge neutrality point shifts as the values of the gate-voltages are varied – this is an outcome of

the modulation of the charge carrier density $n$ and the displacement field $D$ perpendicular to the device [3].

## S2. WEAK ANTI-LOCALIZATION FITTING USING MF EQUATION IN BLG/WSE$_2$

BLG intrinsically shows weak-localization (WL) correction to the Drude conductance because of a berry phase of $2\pi$. An induced strong SOC can modify the quantum interference phenomenon that will give a negative magnetoconductance at low magnetic fields leading to weak anti-localization. In Fig. S2(a), (b) and (c) the change in conductance are plotted versus $B$ for the points J ($D \sim -0.33$ V/nm), I ($D \sim -0.35$ V/nm) and L ($D \sim -0.39$ V/nm) respectively. These points are marked in Fig. 2 of the main text. In the main text we have shown the fits to the data using the model of S. Ilić et al [4]. Here, for comparison, we fit the data using the MF equation [5]:

$$\Delta\sigma(B) = \sigma(B) - \sigma(0) = -\frac{e^2}{2\pi h}\left[F\left(\frac{\tau_B^{-1}}{\tau_\phi^{-1}}\right) - F\left(\frac{\tau_B^{-1}}{\tau_\phi^{-1} + 2\tau_{asy}^{-1}}\right) - 2F\left(\frac{\tau_B^{-1}}{\tau_\phi^{-1} + \tau_{asy}^{-1} + \tau_{sym}^{-1}}\right)\right] \quad (1)$$

where, $F(x) = ln(x) + \psi(1/2 + 1/x)$ and $\psi(x)$ the digamma function, $\tau_B^{-1} = 4e\tilde{D}B/\hbar$, $\tilde{D}$ is the diffusion constant evaluated using the relation by $\tilde{D} = \sigma\pi\hbar^2/2m^*e^2$ [6]. $\sigma$ is the average conductivity in each region in the $n - \tilde{D}$ space, $m^* = 0.033m_o$ and $m_o$ is mass of free electron, $\tau_\phi^{-1}$ is the coherent scattering rate, $\tau_{asy}^{-1}$ is the scattering rate because of the spin-orbit coupling term which breaks the out of plane inversion symmetry. $\tau_{sym}^{-1}$ is the scattering rate due to SOC that preserves the out of plane inversion symmetry. The fits are shown in Fig. S2 by red solid lines. The extracted time scales are listed in table. I.

For the WAL fit one can observe the two time scales $\tau_{asy}$ and $\tau_{so}$ are coming out to be same for all the fits. Since $\tau_{so}^{-1} = \tau_{asy}^{-1} + \tau_{sym}^{-1}$, this implies that the $\tau_{sym}^{-1}$ is vanishingly small. It should be noted here that vanishingly symmetric rate statement is based on the fit parameters obtained from the MF equation, which is an unphysical result. In fact, the reduced equation (2) of main manuscript used to fit the WAL data is obtained in the limit $\Delta_{so} > \tau_{iv}^{-1}$. Since the valley Zeeman coupling is the major source of symmetric rate of spin orbit coupling, it is the dominant scattering term for spin relaxation. Fit for the WL data are shown in Fig. S2 (d), (e) and (f) for D ($D \sim 0.24$ V/nm), K ($D \sim 0.27$ V/nm)

| location | $D$ (V/nm) | $\tau_\phi$ (ps) | $\tau_{asy}$ (ps) | $\tau_{so}$ (ps) | Regime |
|---|---|---|---|---|---|
| J | -0.33 | 22.88 | 0.25 | 0.25 | WAL |
| I | -0.35 | 14.14 | 0.23 | 0.23 | WAL |
| L | -0.39 | 11.50 | 0.29 | 0.29 | WAL |
| location | $D$ (V/nm) | $\tau_\phi$ (ps) | $\tau_{asy}$ (ps) | $\tau_{so}$ (ms) | Regime |
| D | 0.24 | 9.32 | 6.54 | 0.43 | WL |
| K | 0.27 | 13.79 | 0.31 | 0.16 | WL |
| C | 0.30 | 11.18 | 11.33 | 0.077 | WL |

Table I: Extracted time scales from fitting the magneto-conductance data using MF equation. The top table shows the time scales extracted for WAL data. The relevant time scales are $\tau_\phi$, $\tau_{asy}$ and $\tau_{so}$. The bottom table shows the time scales extracted for the WL data.

and C ($D \sim 0.30$ V/nm) points, respectively. The extracted phase coherence time $\tau_\phi$ came nearly the same as found in the main text (and mentioned in table. II). On the other hand, $\tau_{so}$ was found to be of the order of $\sim$ few 100 $\mu$s, which is nonphysical because of the criteria $\tau_{so}^{-1} = \tau_{asy}^{-1} + \tau_{sym}^{-1}$. This is one of the reasons behind the claim that MF equation is not sufficient to capture all the regimes, because it does not account for the the inter-valley scattering time $\tau_{iv}$, which is essential to discuss the WL regime.

## S3. FITS TO MAGNETO-CONDUCTANCE DATA

As seen in the previous section, the MF equation does not account for spin split bands. So, we used the models developed by S. Ilic et al. [4] that takes into account the scattering mechanisms of the spin split bands in our device. The relevant scattering time scales for the case of WAL are $\tau_\phi$ and $\tau_{asy}$, where $\tau_{asy}^{-1}$ dominates; and those for the WL are $\tau_\phi$ and $\tau_{iv}$, with the $\tau_{iv}^{-1}$ dominating in this case. The parameters extracted from the fit are discussed in the main manuscript and tabulated in table II. The only spin dependent scattering time relevant in the WAL case is $\tau_{asy}$, as pointed out in the case of MF equation fits.

5

| location | $D$ (V/nm) | $\tau_\phi$ (ps) | $\tau_{asy}$ (ps) | Regime | location | $D$ (V/nm) | $\tau_\phi$ (ps) | $\tau_{iv}$ (ps) | Regime |
|---|---|---|---|---|---|---|---|---|---|
| J | -0.33 | 13.3 | 1.16 | WAL | D | 0.24 | 9.88 | 1.59 | WL |
| I | -0.35 | 8.82 | 0.72 | WAL | K | 0.27 | 13.35 | 0.92 | WL |
| L | -0.39 | 7.76 | 1.25 | WAL | C | 0.30 | 11.2 | 1.91 | WL |

Table II: Extracted time scales from fitting the magneto-conductance data. The left table shows the time scales extracted for WAL data using Eqn.(2) given in main text. The relevant time scales were found to be $\tau_\phi$ and $\tau_{asy}$. The right table shows the time scales extracted for the WL data using Eqn.(1) given in main text. In this regime, spin scattering is completely washed out and the quantum correction is controlled by inter-valley scattering time $\tau_{iv}$.

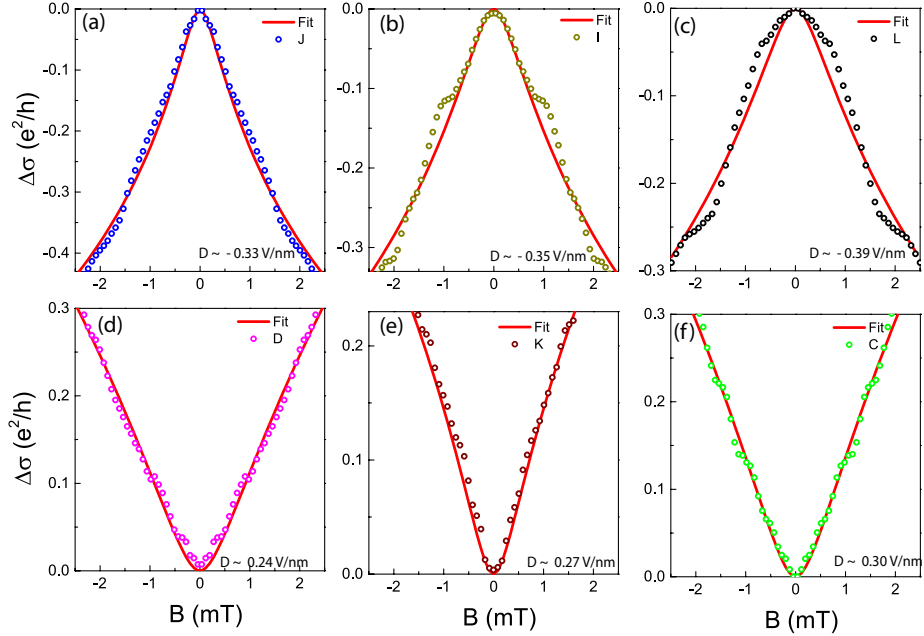

Figure S2: Plot of $\Delta\sigma$ versus magnetic field $B$ for (a) $D = -0.33$ V/nm, (b) $D = -0.35$ V/nm, (c) $D = -0.39$ V/nm, (d) $D = 0.24$ V/nm, (e) $D = 0.27$ V/nm and (f) $D = 0.30$ V/nm respectively. The solid red lines in each panel is the fit to the data using Eqn. 1.


			
\end{document}